\documentclass[mathleft
]{an}
\usepackage{graphicx}
\usepackage{times}
\usepackage[authoryear]{natbib}
\bibpunct{(}{)}{;}{a}{}{,}
\usepackage{times}
\usepackage{amsmath}                  
\usepackage{amssymb}                  
\usepackage{graphicx}
\usepackage{units}
\usepackage{dcolumn}
\usepackage{longtable}
\usepackage{lscape}
\usepackage{booktabs}
\usepackage{aas_macros}
\usepackage{hyperref}

\makeatletter
\newcolumntype{d}[1]{>{\DC@{.}{{.}}{#1}}c<{\DC@end}}
\newcolumntype{o}[1]{>{\DC@{p}{\pm}{#1}}c<{\DC@end}}
\makeatother

\let\orgautoref\autoref

\renewcommand{\autoref}
        {\def\equationautorefname{equation}%
         \def\figureautorefname{Fig.}%
         \def\subfigureautorefname{Fig.}%
         \def\partautorefname{part}%
         \def\chapterautorefname{chapter}%
         \def\sectionautorefname{section}%
         \def\subsectionautorefname{section}%
         \def\subsubsectionautorefname{section}%
         \def\appendixautorefname{appendix}%
         \def\Itemautorefname{item}%
         \def\tableautorefname{Table}%
         \def\lstlistingautorefname{Listing}%
         \orgautoref}
    
\makeatletter

\newcommand{\dsunheute}{d_{\odot,today}}
\newcommand{\dsunSN}{d_{\odot,SN}}

\begin{document}

\Pagespan{789}{}
\Yearpublication{2006}%
\Yearsubmission{2005}%
\Month{11}%
\Volume{999}%
\Issue{88}%

\title{New Radial Velocities for 30 Candidate Runaway
Stars and a Possible Binary Supernova Origin for HIP 9470 and PSR\,J0152$-$1637}

\author{N. Tetzlaff\inst{1}\fnmsep\thanks{Corresponding author:
  \email{nina@astro.uni-jena.de}\newline}
\and  G. Torres\inst{2}
\and  A. Bieryla\inst{2}
\and  R. Neuh\"auser\inst{1}
}
\titlerunning{New Radial Velocities and The Nature of 30 Candidate Runaway Stars}
\authorrunning{N. Tetzlaff et al.}
\institute{
Astrophysikalisches Institut und Universit\"ats-Sternwarte Jena, Schillerg\"asschen 2-3, 07745 Jena, Germany
\and 
Harvard-Smithsonian Center for Astrophysics, 60 Garden St., Cambridge, MA 02138, USA}

\received{.}
\accepted{.}
\publonline{.}


\abstract{%
  We report new radial velocity measurements for 30 candidate runaway stars. We revise their age estimates and compute their past trajectories in the Galaxy in order to determine their birthplaces. We find that seven of the stars could be younger than $\sim$100 Myr, and for five of them we identify multiple young clusters and associations in which they may have formed. For the youngest star in the sample, HIP 9470, we suggest a possible ejection scenario in a supernova event, and also that it may be associated with the young pulsar PSR J0152$-$1637. Our spectroscopic observations reveal seven of the stars in the sample of 30 to be previously unknown spectroscopic binaries. Orbital solutions for four of them are reported here as well.
}

\maketitle

\section{Introduction}
The first studies of high velocity stars were carried out by A.\ Blaauw and collaborators as early as the 1950's \citep{1952BAN....11..414B,1954ApJ...119..625B}.  \citet{1961BAN....15..265B} examined a number of O- and B-type stars and found that many of them had space velocities in excess of 40\,km/s. He referred to these objects as ``runaway stars", a term that has been used ever since.

Runaway stars are mainly produced via two mechanisms. The binary supernova scenario (BSS) was first proposed by \citet{1961BAN....15..265B}. It is also related to the formation of high velocity neutron stars. The runaway and the neutron star are the products of a supernova within a binary system. The velocity of the former secondary (the runaway star) may be as large as its original orbital velocity \citep{1998AA...330.1047T}. The kinematic age of a BSS runaway star is smaller than the age of its parent association since the primary star needed some time to evolve and experience a supernova. In the alternate dynamical ejection scenario (DES) that was suggested by \citet{1967BOTT....4...86P}, stars are ejected from young, dense clusters via gravitational interactions between the cluster members. The (kinematic) age of a DES runaway star should be comparable to the age of its parent association since gravitational interactions are most efficient soon after formation.

To investigate the frequency of runaway stars among all stars in the Galaxy, \citet{1991AJ....102..333S} proposed fitting the velocity distribution with two Maxwellians, one representing normal Population I stars, the second one explaining the group of runaway stars. While many studies aiming at identifying runaway stars were based on space velocities (e.\,g. \citealt{1961BAN....15..265B}), tangential velocities (e.\,g. \citealt{1998A&A...331..949M}) or radial velocities (e.\,g. \citealt{1974RMxAA...1..211C}) alone, we recently constructed a unified catalogue of young runaway stars by evaluating all these criteria as well as the direction of motion compared to its neighbourhood \citep{2011MNRAS.410..190T}. 

Although absolute 3D velocities are not essential to identify runaway stars, they are necessary for constructing stellar orbits. Past stellar trajectories can be used to identify the parent association or cluster of a star and its runaway origin. Some runaway stars have been proposed as former companion candidates to the progenitors of neutron stars \citep{2001A&A...365...49H,2011MNRAS.417..617T,2012PASA...29...98T,2013PhD...Nina}. Such an association may explain the spectral properties of a runaway star, such as a high Helium abundance or high rotational velocity or even enhanced $\alpha$ element abundances as supernova debris \citep{1993ASPC...35..207B,2008ApJ...684L.103P}. \\

In this paper we investigate the nature of 30 of the candidate runaway stars proposed by \citet{2011MNRAS.410..190T}. Kinematic information for these objects was previously incomplete because of missing radial velocities ($v_r$). Here we present multiple measurements of $v_r$ for these stars, and in addition we revise their age estimates from the work mentioned above. With this information we construct and trace back the Galactic orbits for the young stars in the sample to investigate their origin and possible connection with young clusters and neutron stars. Seven of the targets are revealed to be spectroscopic binaries, and orbits for four of them are reported here.

\section{Spectroscopy and Radial Velocities}\label{sec:spectroscopy}

Spectroscopic observations of our 30 targets were gathered using the Tillinghast Reflector Echelle Spectrograph (TRES, \citealt{Furesz08}) on the 1.5\,m Tillinghast reflector at the F.\ L.\ Whipple Observatory (Mount Hopkins, Arizona, USA), from 2011 April to 2013 October. Individual stars were observed for periods ranging from $\sim$1 yr to a little over 2 yr. The spectra cover the wavelength range  $\sim$3900--8900\,\AA, and were taken with a typical resolving power of $R \sim 44,\!000$ with the medium fiber of the instrument. The signal-to-noise ratios for the individual exposures range between 23 and 230 per resolution element of $\unit[6{.}8]{km/s}$, and refer to the region of the Mg I b triplet ($\sim$5200\,\AA).  All spectra were reduced and extracted using standard procedures as described by \citet{Buchhave:10a} and \citet{Buchhave:10b}.

Radial velocities were obtained by cross-correlation against a synthetic template taken from a large library computed by John Laird, based on model atmospheres by R.\ L.\ Kurucz and a line list developed by Jon Morse \citep[see][]{2012Natur.486..375B}. These templates cover a 300\,\AA\ window centred at 5200\,\AA, which generally contains most of the velocity information, and have a spacing of 250\,K in temperature ($T_{\rm eff}$), $0{.}5$ dex in surface gravity ($\log g$) and metallicity ([Fe/H]), and a variable step in the projected rotational velocity ($v \sin i$).  The optimum template for each star was determined by cross-correlating all observations against the entire library of templates, and choosing the one giving the highest correlation coefficient averaged over all exposures \citep{2002AJ....123.1701T}. The best-matched template provides an estimate of the stellar parameters, although their accuracy is limited by degeneracies that are present among $T_{\rm eff}$, $\log g$, and [Fe/H]. The template parameters that affect the velocities the most are $T_{\rm eff}$ and $v \sin i$. To suppress correlations, in a second iteration we held the metallicity fixed at the solar value and $\log g$ at values determined with the help of stellar evolution models as described in the next section. We then redetermined $T_{\rm eff}$ and $v \sin i$, and recomputed the velocities.

The zero point of our velocity system was monitored by observing standard stars each night with the same instrumental setup. Individual heliocentric velocities for each of our candidate runaway stars are reported in \autoref{tab:indivrv} (the complete table is available as supplementary material).  In \autoref{tab:rvs} we report the average $v_r$ of each star, along with the template parameters and other pertinent information.
\begin{table}
\centering
\caption{Individual radial velocities (with 1$\sigma$ errors) for each of the 30 stars in our sample. Only the measurements for HIP 8414 are shown here. The full table is available as Supporting Information with this article. Here only the measurements for HIP 8414 are shown (this star happens to have variable radial velocity; see \autoref{fig:SB}).}\label{tab:indivrv}
\begin{tabular}{c d{5.4} o{6.4}}
\toprule
HIP				& \multicolumn{1}{c}{HJD-2,400,000}		& \multicolumn{1}{c}{$v_r$}\\\midrule
8414  &    55757.9805 & -29{.}88p0{.}60\\
8414  &    55824.9922 &  -2{.}13p0{.}57\\
8414  &    55840.9297 & -21{.}22p0{.}53\\
8414  &    55884.7539 & -34{.}63p0{.}78\\
8414  &    55905.6914 & -34{.}76p0{.}64\\
8414  &    55960.6094 &   8{.}42p0{.}60\\
8414  &    55983.5820 & -17{.}31p0{.}62\\
8414  &    56231.7500 &   2{.}51p0{.}47\\
8414  &    56253.6953 & -34{.}13p0{.}56\\
8414  &    56316.5703 & -33{.}71p0{.}57\\
8414  &    56496.9492 & -12{.}87p0{.}64\\
8414  &    56549.8203 & -32{.}10p0{.}69\\
8414  &    56550.8438 &  -8{.}26p0{.}58\\
8414  &    56559.9063 & -28{.}48p0{.}59\\
8414  &    56574.9102 &  14{.}45p0{.}76\\
8414  &    56592.7305 & -16{.}79p0{.}59\\
\bottomrule
\end{tabular}
\end{table}
\begin{table*}
\centering
\caption{Mean radial velocities $v_r$ of 30 stars (column 5). The errors represent 1$\sigma$ uncertainties.\newline Columns 2 to 4 list the stellar properties: visual magnitude $\mathbf{V}$, effective temperature ($T_{\rm eff}$ in K), surface gravity in standard (cgs) units ($\log g$) and projected rotational velocity ($v\sin i$ in km/s). In columns 6 and 7 we give the number of spectra ($N$) and the time (in days) spanned by the spectroscopic observations for each star, respectively. The last column contains information on multiplicity (SB1 for single-lined spectroscopic binary).}\label{tab:rvs}
\begin{tabular}{r d{1.2} o{4.3} d{1.2} o{3.2} o{6.4} >{$}r<{$} r p{5.5cm}}
\toprule
HIP					& \multicolumn{1}{c}{$V$} & \multicolumn{1}{c}{$T_{\rm eff}$}	& \log g	& \multicolumn{1}{c}{$v\sin i$} & \multicolumn{1}{c}{$v_r$}	& 	N		&	span	&	Notes \\\midrule
   8414 &   8.50& 6600p100  &   4.23  &   42p4  &    -13{.}06p0{.}22 &   16  &   835  & SB1, $P=\unit[1{.}74]{d}$, additional visual companion at 1{.}7'' \citep{2007AA...464..641L}, implying a triple system \\
   9470 &   6.54& 5000p100  &   1.54  &    8p2  &    -11{.}91p0{.}10 &   12  &   749  & SB1, $P=\unit[558]{d}$\\
  10784 &   6.88& 4600p100  &   1.92  &    4p1  &    +34{.}85p0{.}05 &    7  &   400  & \\
  18549 &   8.51& 8000p250  &   4.16  &  151p15 &      +3{.}97p1{.}35&     6 &    351 & \\
  27802 &   10.39& 4700p100  &   4.19  &    2p1  &    +14{.}75p0{.}06 &   11  &   436  &\\
  39228 &   10.06& 5600p100  &   4.56  &    1p1  &     -4{.}51p0{.}05 &    8  &   396  &Secondary in double system with HIP 39226 \\ 
  42331 &   7.60& 4050p100  &   1.27  &    4p1  &     +3{.}83p{0{.}37} &     17  &   774 &  Long-term drift\\
  43043 &   7.99& 7400p100  &   4.20  &   56p4  &    +25{.}11p0{.}09 &    9  &   396  &\\
  49820 &   10.36& 7150p100  &   4.29  &   51p4  &     +9{.}12p0{.}13 &   13  &   632 & \\
  60065 &   10.36& 7350p150  &   4.32  &  152p15 &     -15{.}58p0{.}79&     9 &    388&  \\
  70995 &   10.60& 5900p100  &   4.56  &    2p1  &     -3{.}39p0{.}07 &    9  &   388 & \\
  82163 &   7.91& 4200p100  &   1.54  &    4p1  &    -64{.}77p{1{.}12} &     16  &   773  & Long-term drift\\
  82304 &   7.35& 3850p100  &   1.16  &    6p1  &    -27{.}41p0{.}04 &     9  &   416  &\\
  83627 &   8.46& 7700p100  &   4.21  &   11p2  &    -20{.}56p0{.}11 &   10   &  347  &\\
  84385 &   7.12& 3650p100  &   1.03  &    7p2  &    +15{.}00p0{.}17 &   19   &  768  &\\
  85271 &   7.74& 3800p100  &   1.31  &    6p1  &    -53{.}01p0{.}06 &   14   &  768  &\\
  91545 &   8.32& 7350p150  &   4.25  &  159p15 &     -43{.}99p0{.}89&     7  &   360 & \\
  96045 &   7.61& 4100p100  &   1.37  &    5p1  &    -52{.}12p0{.}05 &    8   &  530 & \\
  98019 &   8.16& 4500p100  &   1.79  &    6p1  &     -4{.}22p0{.}03 &    7   &  381  &\\
  98443 &   6.98& 4050p100  &   1.43  &    6p1  &    -42{.}54p{1{.}37}  &    15  &   731 &  Long-term drift, primary in triple system \citep{1987SvA....31..220A}\\
  101219&   7.50& 3700p100 &    0.98 &     4p1 &      -5{.}00p0{.}06 &   15  &   895  & SB1, $P=\unit[483]{d}$ \\
  101320&   7.68& 4900p100 &    1.95 &     5p1 &     -18{.}74p0{.}01 &    7  &   377  &\\
  102000&   9.14& 9150p100 &    4.28 &    80p4 &      +2{.}76p0{.}47 &   14  &   751  &\\
  103533&   8.61& 5500p150 &    3.02 &   154p20&      +26{.}67p4{.}03&    12 &    751 & Secondary in double system with HIP 103537 \\
  104581&   10.09& 5300p100 &    4.61 &     2p1 &     -21{.}54p0{.}06 &    6  &   514  & Primary in double system with K3 type star \\
  104608&   11.01& 7250p100 &    4.17 &    50p4 &      +3{.}23p0{.}25 &    6  &   509  &\\
  106291&   9.25& 7250p100 &    4.32 &    20p3 &      +5{.}21p0{.}03 &    6  &   514  &\\
  111607&   8.06& 4350p100 &    1.76 &     3p1 &     +21{.}64p0{.}05 &    7  &   718  &\\
  113787&   6.56& 4800p100 &    2.04 &     5p1 &      +4{.}40p0{.}02 &   13  &   589  & SB1, $P=\unit[467]{d}$\\
  117998&   8.30& 7450p100 &    4.23 &    94p5 &     +12{.}25p0{.}50 &    8  &   414  &\\\bottomrule
\multicolumn{9}{p{0.9\textwidth}}{Note: The $\log g$ values were estimated from stellar evolution models
based on preliminary temperature estimates, and were held fixed (along
with an assumed solar metallicity) to infer the final temperatures reported here (see text). The uncertainties reported for the velocities are the standard error of the mean.}\\
\end{tabular}
\end{table*}

\subsection{Binary systems}

\begin{figure}
\centering
\includegraphics*[width = 0.47\textwidth,viewport = 55 55 555 750]{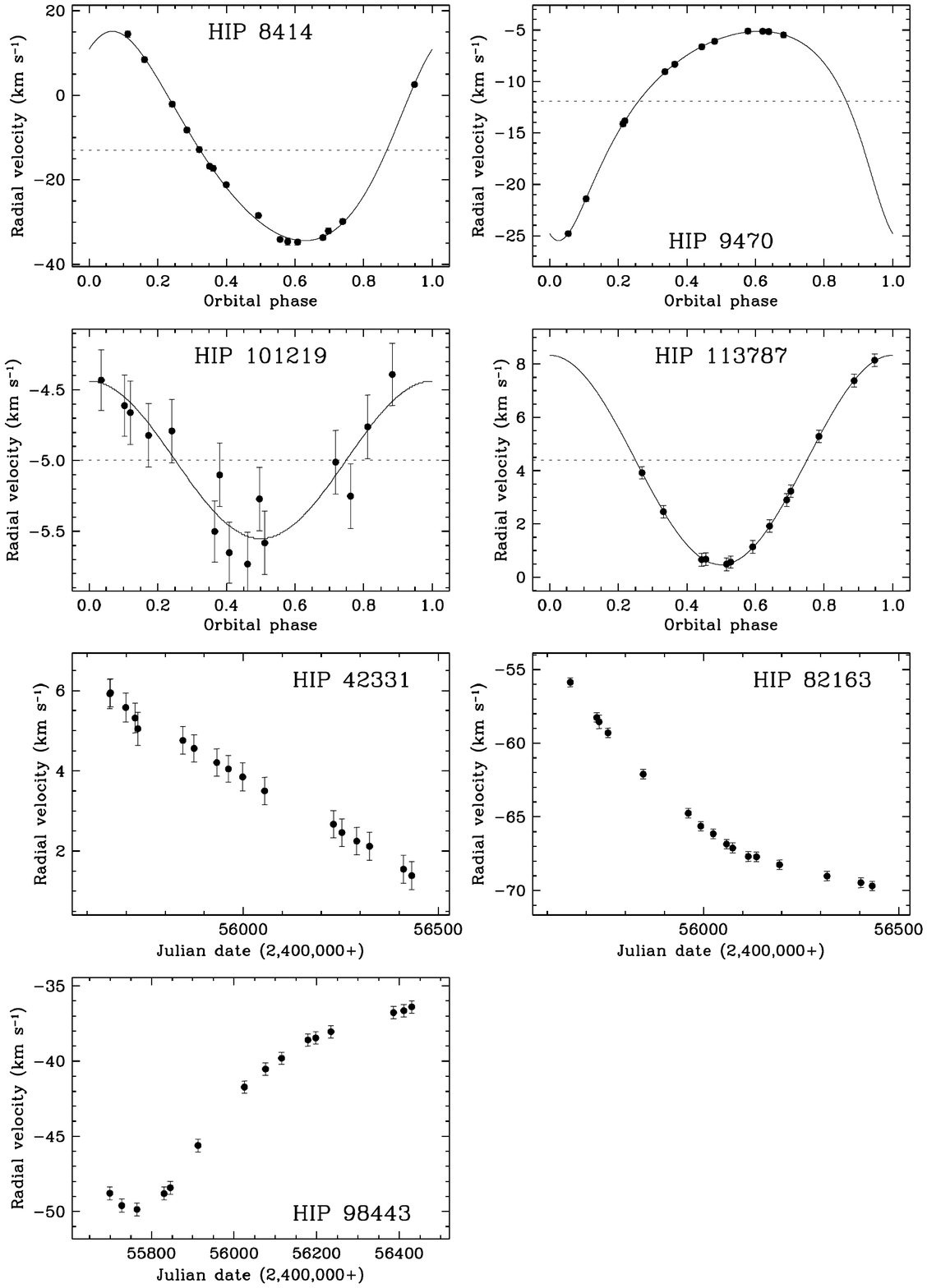}
\caption{Orbital solutions for four binary stars and time histories of three objects with long-term trends.}\label{fig:SB}
\end{figure}

Seven of our targets have variable radial velocity and are spectroscopic binaries. For four of them we gathered sufficient observations to solve for the orbital elements, which are presented in \autoref{tab:SB}. In two cases (HIP\,101219 and HIP\,113787) the eccentricities turned out not to be significant; the solutions presented for those stars assume the orbits to be circular.  The velocities listed for these binaries in \autoref{tab:rvs} correspond to the systemic velocity. The other three variables show only long-term drifts over the time span of our observations, and their orbital periods are likely to be several years. Therefore, the mean velocities listed for these stars in \autoref{tab:rvs} do not necessarily represent the true centre-of-mass velocity. The orbital solutions are shown graphically in \autoref{fig:SB}, along with the time histories of the three objects with long-term trends.

Three of the four stars with spectroscopic orbits (HIP\,9470, HIP\,101219, and HIP\,113787), as well as the three with long-term trends, are slowly rotating G or K giants (see below) and their orbital periods are long. The other binary with a spectroscopic orbit  (HIP\,8414) has a short period, and the rapidly rotating primary star is an F dwarf. Its measured rotation ($v \sin i = \unit[42]{km/s}$) is consistent with the star being pseudo-synchronized with motion in the slightly eccentric orbit \citep{1981A&A....99..126H}.
\begin{table*}
\centering
\caption{Orbital solutions for four binary stars.\newline
$P$ -- orbital period, $\gamma$ -- systemic velocity, $K$ -- velocity semi-amplitude of the primary, $e$ -- eccentricity, $\omega$ -- longitude of periastron, $T$ -- time of periastron passage for systems with eccentric orbits or time of maximum velocity for systems with
circular orbits, $a_1\sin i$ -- projected semimajor axis of the primary star, $f\left(M\right)$ -- mass function, $M_2\sin i$ -- minimum mass of secondary, $\sigma$ -- root-mean-square residual from the fit.}\label{tab:SB}
\begin{tabular}{@{} l >{$}c<{$} >{$}c<{$} >{$}c<{$} >{$}c<{$} >{$}c<{$} >{$}c<{$} >{$}c<{$} @{}}
\toprule
Element & $HIP 8414$ & $HIP 9470$ & $HIP 101219$ & $HIP 113787$ \\
\midrule
$P$ [d]   &          1{.}740469 \pm 0{.}000020  &   556{.}5 \pm 1{.}5    &       545 \pm 23       &        467{.}12 \pm 0{.}94\\
$\gamma$ [km/s] &         -13{.}06 \pm 0{.}22  &     -11{.}912 \pm 0{.}099  &    -4{.}995 \pm 0{.}055    &         4{.}402 \pm 0{.}016\\
$K$ [km/s] &              24{.}77 \pm 0{.}43   &   10{.}174 \pm 0{.}088     &   0{.}56 \pm 0{.}08    &          3{.}932 \pm 0{.}022\\
$e$ &              0{.}165 \pm 0{.}018  &     0{.}3568 \pm 0{.}0064   &      0\mbox{ (fixed)}        &           0\mbox{ (fixed)}\\   
$\omega$ [deg] &               326{.}0 \pm 3{.}7   &       159{.}06 \pm 0{.}83      &         \ldots &  \ldots\\     
$T$ [HJD] &   2456205{.}734 \pm 0{.}017  &   2456194{.}6 \pm 2{.}4  &   2456178 \pm 13      &     2455985{.}20 \pm 0{.}44\\
$a_1\sin i$ [Gm] &     0{.}5847 \pm 0{.}0090   &    72{.}73 \pm 0{.}74   &      4{.}17 \pm 0{.}64     &         25{.}26 \pm 0{.}13 \\
$f\left(M\right)$ [M$_\odot$] &          \left(2{.}63 \pm 0{.}12\right)\cdot10^{-3}  &   \left(4{.}95 \pm 0{.}15\right)\cdot10^{-2}  &    \left(9{.}7  \pm 4{.}1 \right)\cdot10^{-6} &   \left(2{.}943 \pm 0{.}048\right)\cdot10^{-3}\\
$M_2\sin i$ [$\left(M_1+M_2\right)^{2/3}\mathrm[M_\odot]$] &      0{.}1380 \pm 0{.}0021  &    0{.}3672 \pm 0{.}0036   &  0{.}0213 \pm 0{.}0030  &        0{.}14330 \pm 0{.}00078\\
number of observations &                 16        &             12           &         15          &                13\\
span of observations [d] &                 834{.}8       &           748{.}9    &             894{.}6    &                   588{.}6\\
$\sigma$ [km/s] &                  0{.}81         &          0{.}06       &           0{.}20          &              0{.}05\\\bottomrule
\end{tabular}
\end{table*}

\section{Past Trajectories}

In order to construct the past orbits of our sample stars, stellar ages are required. In our previous study \citep{2011MNRAS.410..190T} we already reported age estimates by comparison with evolutionary models in order to focus on the younger runaway candidates, as past orbits become too uncertain for older objects. The temperatures of the stars used to place them on the H-R diagram were based on published spectral types, and the conversion table of \citet{Schmidt-Kaler1982}. However, because of variations in chemical composition the zero-age main sequence (ZAMS) has some spread at a given effective temperature, which could not be resolved for the objects in our earlier study because we lacked estimates of the metallicity. To be conservative and not miss potentially young objects, late-type stars that were found to fall below the solar-metallicity ZAMS in the H-R diagram were shifted upwards in luminosity onto the model ZAMS. This most likely resulted in underestimated ages for those stars.

\begin{figure}
\centering
\includegraphics[width = 0.45\textwidth,viewport = 40 205 540 600]{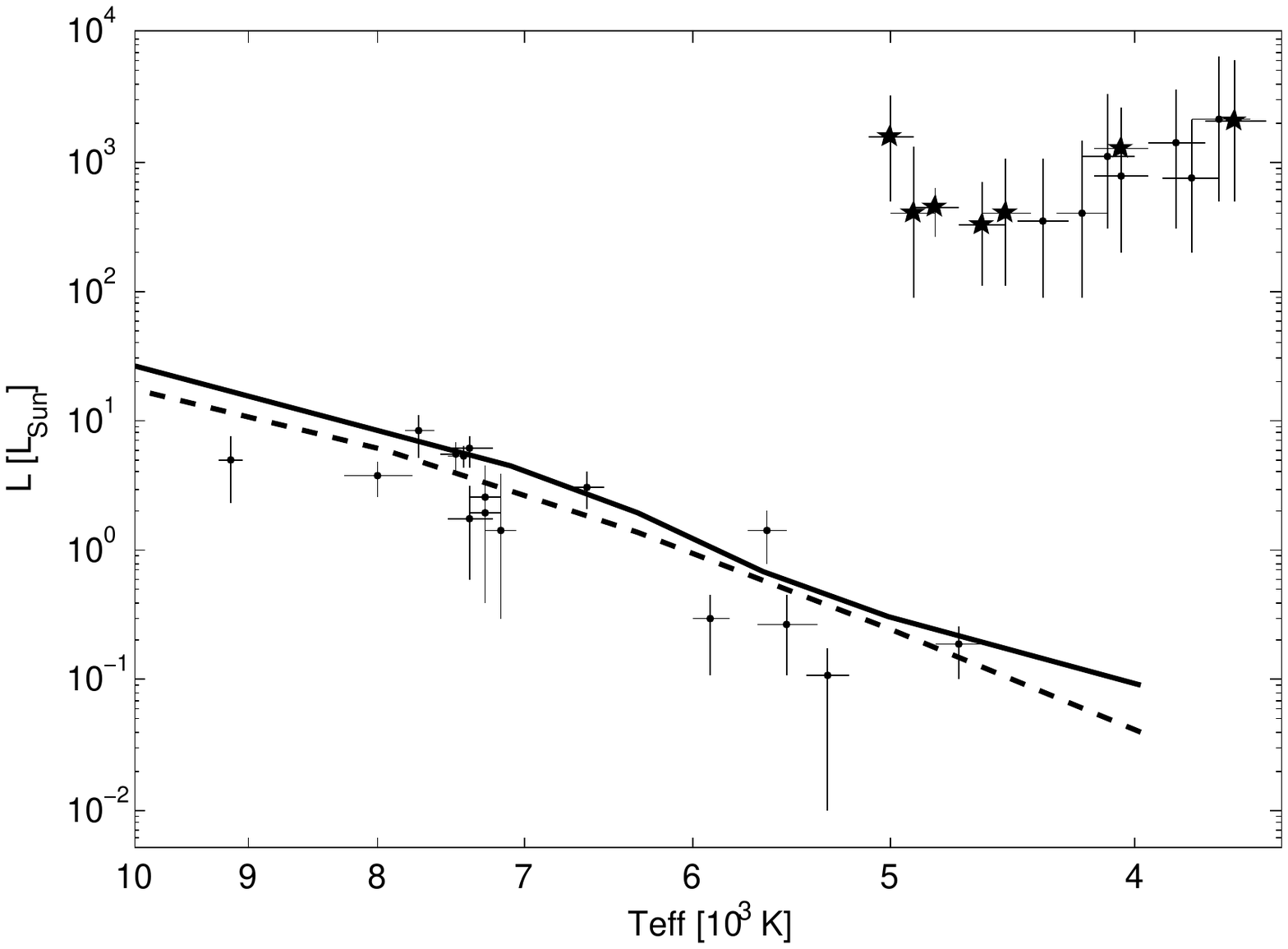}
\caption{The 30 stars in our sample shown on the H-R diagram. Objects indicated by a star are possibly young (up to $\sim\unit[100]{Myr}$) stars. The solid and dashed lines indicate the mean model ZAMS for solar metallicity and $\mathrm{[Fe/H]}~=~-0{.}4$, respectively.}\label{fig:HRD}
\end{figure}

For the 30 stars in the present sample we now have available the spectroscopic temperatures determined by cross-correlation. To compute stellar luminosities $L$, we determined the distances from their HIPPARCOS parallaxes \citep{2007AA...474..653V} using equation 2.11 of \citet{2013MNRAS.436.1343F}. This prescription yields results that are on average less affected by the bias towards larger distances arising from the parallax errors and the reciprocal relation between distance and parallax\footnote{The distance formula by \citet{2013MNRAS.436.1343F} is strictly valid for parallax errors below 20 per cent. Although many of the stars in our sample exceed this limit, we have chosen to use the same expression because numerical simulations of a sample of 20,000 stars show that the resulting distances are still closer to the true distances than the nominal values. This procedure is also preferable to other approaches such as that of \citep{1996MNRAS.281..211S}, as shown by the same simulations.}. Two of the stars, HIP 39228 and HIP 103533, are the secondaries in visual pairs, and their nominal parallaxes have very large uncertainties caused by the presence of the brighter primaries, and may be biased for the same reason. For those stars we adopted the catalogue values listed for the primaries. We further used the $V$ magnitudes as reported in the Tycho-2 catalog \citep{2000A&A...355L..27H} converted to the Johnson system, and an estimate of the extinction based on the $B-V$ indices (converted from the Tycho-2 measurements) and standard colours for main sequence and giant stars (see \citealt{2011MNRAS.410..190T} and \citealt{2013PhD...Nina}). Approximately half of our stars are found to be evolved.  Ages and masses were inferred by comparison with several series of stellar evolution calculations (\citealt{1992AAS...96..269S,1993A&AS...98..523S,2004ApJ...612..168P,2008AA...484..815B,2009AA...508..355B}\footnote{Isochrones from \citet{1992AAS...96..269S} and \citet{1993A&AS...98..523S} were obtained from the database http://webast.ast.obs-mip.fr/equipe/stellar/ \citep{2001A&A...366..538L}. For \citet{2004ApJ...612..168P}, see also the BaSTI web tools (http://albione.oa-teramo.inaf.it/).}), accounting for all uncertainties by utilizing a Monte Carlo method. The corresponding $\log g$ values were used to help establish the spectroscopic temperatures, as described earlier. However, the limitation due to the unknown  metallicities remains, as we cannot determine [Fe/H] reliably from our spectroscopic material due to the strong correlations with $T_{\rm eff}$ and $\log g$ mentioned in \autoref{sec:spectroscopy}.  Therefore, we do not determine ages and masses for nine stars that fall below the solar-metallicity ZAMS (\autoref{fig:HRD}); those objects are likely old in any case, which prevents a reliable determination of their Galactic orbits. The masses and ages determined for the remaining 21 stars are listed in \autoref{tab:massage}. We note that among these 21 stars there are 14 stars that are located in the upper right part of the HR diagram and, hence, are possibly evolve stars. Some of them are probably already old.\\
\begin{table}
\centering
\caption{Evolutionary ages $\tau_\star$ and masses $m_\star$ for 21 stars. The errors include the uncertainty on the parallax, $V$ magnitude, $B-V$ colour, $T_{\rm eff}$, the spread between different evolutionary models as well as a lower metallicity of $\mathrm{[Fe/H]}=-0{.}4$. The nominal values for mass and age represent the respective median for solar metallicity. Stars that are marked with an asterisk (*) could clearly be younger than $\unit[100]{Myr}$, the other stars are most probably a few hundred up to a few thousand Myr old. Note that we did not consider HIP 82304, HIP 83627, HIP 111607 and HIP 117998 young stars. For the first three of these stars, median ages of $\sim\unit[600-800]{Myr}$ suggest that the stars are not young and the lower age bound does not clearly fall below our $\unit[100]{Myr}$ limit. For HIP 117998 the lower age bound is relatively small suggesting that the star is older than $\unit[100]{Myr}$.}\label{tab:massage}
\setlength\extrarowheight{2pt}
\begin{tabular}{r >{$}c<{$} >{$}c<{$}}
\toprule
HIP			&	\tau_\star			& m_\star \\
				&	$[Myr]$ 				& $[M$_\odot$]$ \\	\midrule
8414    & 2000^{+2000}_{-1300}						& 1{.}3^{+0{.}1}_{-0{.}3} \\
9470$^*$    & 65^{+91}_{-20}									& 6{.}0^{+2{.}2}_{-2{.}2} \\
10784$^*$   & 168^{+600}_{-70}								& 3{.}9^{+1{.}9}_{-1{.}9} \\
27802   & \sim 14000^\mathrm{a} 					& 0{.}6^{+0{.}3}_{-0{.}1} \\
39228   & \sim 10000^\mathrm{a} 					& 1{.}0^{+0{.}1}_{-0{.}3} \\
42331   & 1000^{+5000}_{-820}							& 3{.}0^{+5{.}9}_{-2{.}6} \\
43043   & 1000^{+2000}_{-540}							&	1{.}5^{+0{.}1}_{-0{.}2} \\
82163   & 570^{+3900}_{-450}							& 3{.}0^{+4{.}6}_{-2{.}1} \\
82304   & 570^{+3500}_{-430}							& 1{.}7^{+3{.}8}_{-1{.}2} \\
83627   & 800^{+1200}_{-700}							& 1{.}7^{+0{.}1}_{-0{.}3} \\
84385$^*$   & 400^{+5500}_{-350}							& 3{.}9^{+1{.}6}_{-3{.}4} \\
85271   & 1000^{+\ldots}_{-800}\ ^\mathrm{b}& 0{.}9^{+2{.}3}_{-0{.}5} \\
91545   & 1000^{+2000}_{-600}							& 1{.}6^{+0{.}1}_{-0{.}3} \\
96045   & 550^{+6500}_{-350}							& 3{.}5^{+6{.}5}_{-2{.}7} \\
98019$^*$   & 150^{+3000}_{-50}								& 4{.}3^{+1{.}6}_{-3{.}8} \\
98443$^*$   & 118^{+5100}_{-50}	  		  			& 4{.}6^{+1{.}6}_{-4{.}1} \\
101219  & 6600^{+7400}_{-6400}						& 2{.}8^{+5{.}6}_{-2{.}5} \\
101320$^*$  & 120^{+14000}_{-70}							& 2{.}9^{+4{.}1}_{-2{.}2} \\
111607  & 570^{+4500}_{-470}  						& 2{.}0^{+4{.}3}_{-1{.}4} \\
113787$^*$  & 168^{+178}_{-79}								& 4{.}3^{+0{.}8}_{-1{.}5} \\
117998  & 200^{+2800}_{-90}								& 1{.}5^{+0{.}1}_{-0{.}1} \\
\bottomrule
\multicolumn{3}{p{0.4\textwidth}}{$^\mathrm{a}$ These stars are too old for precise age estimate with our methods.}\\
\multicolumn{3}{p{0.4\textwidth}}{$^\mathrm{b}$ The upper age limit of HIP 85271 cannot be constrained by our methods.}
\end{tabular}
\end{table}
 
With knowledge of the distance and kinematics (proper motions\footnote{We consistently use proper motions and parallaxes from HIPPARCOS \citep{2007AA...474..653V}.} and our new $v_r$ measurements), we used the fourth-order Runge-Kutta algorithm implemented in \textsc{MATLAB}\textsuperscript{\copyright} to calculate the past stellar trajectories. We adopted an axisymmetric Galactic potential that contains a Miyamoto-Nagai potential for the disk \citep{1975PASJ...27..533M}, a Hernquist potential for the bulge and the inner halo \citep{1990ApJ...356..359H} and a logarithmic potential for the dark halo, see \citet{2001AJ....122.1397H}. For seven stars that could clearly be younger than $\unit[100]{Myr}$ (marked with an asterisk in \autoref{tab:massage}), we compared the past flight paths with those of young associations (see \citealt{2010MNRAS.402.2369T,2012PASA...29...98T,2013PhD...Nina}) and 42 stellar clusters with ages between 40 and $\unit[150]{Myr}$ from \citet{2012AcA....62..281B} for which 3D kinematic data are available. The kinematic properties of the clusters where obtained from \citet{2006AA...446..949D}, \citet{2005A&A...440..403K}, \citet{2009MNRAS.399.2146W} and \citet{2010MNRAS.407.2109V}, or derived as the mean values of members listed in the WEBDA database\footnote{http://www.univie.ac.at/webda/} \citep{2003A&A...410..511M}. Due to the positional uncertainties of the associations/clusters that increase with time because of their velocity dispersion, we calculate the past flight paths only up to $\unit[150]{Myr}$. For the same reason, we did not account for association/cluster expansion or evaporation effects because the positional uncertainties dominate.\\
For five stars, we found multiple possible birth clusters (for flight times below $\unit[150]{Myr}$) that we list in \autoref{tab:birthcl} along with the flight time and the cluster age. In these cases, taking the positional and kinematic uncertainties into account, the runaway star was probably inside the boundaries of the association/cluster within the past $\unit[150]{Myr}$.
\begin{table}
\centering
\caption{Possible birth clusters for five young sample stars along with the flight time of the star. Ages are taken from \citet{2005A&A...440..403K}, \citet{2006AA...446..949D}, \citet{2010MNRAS.407.2109V} and \citet{2012AcA....62..281B}.}\label{tab:birthcl}
\setlength\extrarowheight{2pt}
\begin{tabular}{r l >{$}c<{$} >{$}c<{$}}
\toprule
HIP			&	cluster			& $flight time$  & $cluster age$\\
				&			 				& $[Myr]$ 		 & $[Myr]$  	\\	\midrule
9470		& NGC 103			& 135^{+10}_{-15}& 112-134 \\
				& NGC 1444		& 36^{+57}_{-36} & 79-92 \\
				& NGC 2353    & 91^{+54}_{-28} & 89-126 \\
84385   & NGC 103			& 92^{+10}_{-9}	 & 112-134 \\
				& NGC 436     & 109^{+2}_{-1}  & 84-126 \\
				& NGC 2353    & 50^{+43}_{-37} & 89-126 \\
        & Pismis 16   & 63^{+5}_{-2}   & 52-79 \\
        & NGC 6694    & 67^{+1}_{-1}   & 79-85 \\
        & NGC 1513    & 59^{+115}_{-59}& 79\\
        & NGC 7086    & 98^{+7}_{-4}   & 112-139\\
98019   & NGC 7086    & 105^{+10}_{-15}& 112-139\\
        & NGC 2353    & 100^{+37}_{-27}& 89-126 \\
        & NGC 5617    & 100^{+8}_{-10} & 79-112\\
98443   & BH 92				& 58^{+1}_{-2}   & 56 \\
        & NGC 7086    & 150^{+4}_{-4}  & 112-139\\
113787  & NGC 7788		& 120^{+7}_{-5}	 & 30-158\\
				& NGC 1778    & 134^{+4}_{-3}  & 126-151\\
				& NGC 2186    & 101^{+2}_{-2}  & 55-200\\
\bottomrule
\end{tabular}
\end{table}

The flight time to reach the Galactic plane is smaller than the stellar ages for all cases as half the period of the vertical oscillation is smaller than the stellar age. Therefore, we cannot make statements about whether any of these stars were born in the Galactic plane.\\

We also calculated the past trajectories of young neutron stars in order to find possible encounters with runaway stars. We note, however, that only about 35 per cent of former runaway star/neutron star pairs can be recovered due to the large uncertainties on the observables \citep{2013PhD...Nina}. Owing to the unknown radial velocity of neutron stars, their flight paths can only be traced back up to a maximum of a few Myr (we use $\unit[5]{Myr}$, e.\,g. \citealt{2010MNRAS.402.2369T,2011MNRAS.417..617T,2013PhD...Nina}). This also sets a limit on the maximum age of the runaway star. Assuming that the neutron star is $\unit[5]{Myr}$ old and that the mass of its progenitor was the minimum mass required for a supernova progenitor ($\sim$B3, implying a lifetime of $\sim\unit[30-40]{Myr}$), the only suitable runaway star in our sample is HIP 9470 (allowing for some Myr uncertainty on the progenitor lifetime and the stellar age). 

To search for a neutron star that may potentially be related to HIP 9470, we apply a Monte Carlo method as done previously by \citet{2001A&A...365...49H}, \citet{2009AstL...35..396B}, and by ourselves \citep{2009MNRAS.400L..99T,2010MNRAS.402.2369T,2011MNRAS.417..617T,2012PASA...29...98T,2013MNRAS.435..879T,2014MNRAS.tmp..150T}, to account for all errors on the observables. Among all young nearby neutron stars\footnote{Because of the large uncertainties on the observables of neutron stars and the unknown radial velocity, their flight paths can only be traced back up to a maximum of $\sim\unit[5]{Myr}$. As the spin-down age only gives a rough estimate of the order of magnitude of the true age, neutron stars with spin-down ages up to $\unit[50]{Myr}$ were selected for investigation allowing for an uncertainty on the age of one order of magnitude to be conservative.} in the ATNF pulsar database\footnote{Pulsar database operated by the Australia Telescope National Facility, \citet{2005AJ....129.1993M}, http://www.atnf.csiro.au/research/pulsar/psrcat/.} with available distance and proper motion (105 neutron stars in total, see \citealt{2013PhD...Nina}), we find two candidates for a common origin with HIP 9470: PSR\,J0358+5413 and PSR\,J0152$-$1637. The present neutron star parameters and the position and time of the predicted supernova event are given in \autoref{tab:J0152_J0358_HIP9470}. The kinematic age (flight time) of PSR\,J0358+5413 is $\sim\unit[5]{Myr}$ if HIP 9470 is its former companion. This exceeds the spin-down age by one order of magnitude ($\tau_{sd}=\unit[564]{kyr}$, period $P$ and its derivative $\dot{P}$ from \citealt{2004MNRAS.353.1311H}). \citet{2013MNRAS.430.2281N} determined the kinematic age as $\unit[0{.}5^{+0{.}5}_{-0{.}2}]{Myr}$ assuming that the pulsar was born in the Galactic plane which is in good agreement with $\tau_{sd}$. 
Therefore, we do not consider it likely for the progenitor of PSR J0358+5413 to have been a former companion of HIP 9470. The pulsar was probably born in the Galactic plane half a million years ago.
\begin{table*}
\centering
\caption{Predicted current parameters of PSR\,J0358+5413 and PSR\,J0152$-$1637 and supernova position and time for a possible past encounter with HIP 9470.\newline
Neutron star parameters: heliocentric radial velocity $v_r$, distance $d$ or parallax $\pi$, proper motion $\mu_{\alpha}^*$, $\mu_\delta$, peculiar space velocity $v_{sp}$; Predicted supernova position: distance of the supernova to Earth at the time of the supernova ($\dsunSN$) and as seen today ($\dsunheute$), Galactic coordinates (Galactic longitude $l$, Galactic latitude $b$, J2000.0) as seen from the Earth today; Predicted time of the supernova in the past $\tau$. For the derivation of the parameters we refer the reader to the work of \citealt{2010MNRAS.402.2369T}.}\label{tab:J0152_J0358_HIP9470}
{
\begin{tabular}{l  c c c c}
\toprule
Present-day parameters of  & \multicolumn{2}{c}{PSR\,J0358+5413}&	\multicolumn{2}{c}{PSR\,J0152$-$1637}\\
&								predicted	& measured$^\mathrm{a}$ & predicted & measured$^\mathrm{b}$\\\midrule
$v_r$ [km/s]             	 & $211^{+114}_{-38}$ & --& $134^{+117}_{-59}$	& --\\
$\pi$ [mas] or $d$ [pc] 	& $0{.}7^{+0{.}1}_{-0{.}2}$ & $0{.}91\pm0{.}16$& $833^{+167}_{-119}$ & 510, 790 \\
$\mu_{\alpha}^*$ [mas/yr]  & $9{.}2\pm0{.}2$ & $9{.}20\pm0{.}18$& $2{.}9\pm1{.}1$	& $3{.}1\pm1{.}2$ \\
$\mu_\delta$ [mas/yr]      & $8{.}3\pm0{.}4$ & $8{.}17\pm0{.}39$& $-27{.}5\pm2{.}0$	& $-27\pm2$ \\
$v_{sp}$ [km/s]					 		& $266^{+101}_{-57}$ & --	& $175^{+86}_{-70}$	& -- 	\\\midrule
\multicolumn{5}{c}{Predicted supernova position and time since the explosion}\\\midrule
$\dsunSN$ [pc] &   $453^{+109}_{-52}$& &$464^{+97}_{-82}$ &\\
$\dsunheute$ [pc]  &   $451^{+91}_{-57}$& &$578^{+84}_{-90}$ &\\
$l$ [$\mathrm{^\circ}$]     &  $149{.}4^{+2{.}4}_{-2{.}4}$& &$149{.}9^{+2{.}1}_{-2{.}2}$	&\\
$b$ [$\mathrm{^\circ}$]     &  $-49{.}8^{+1{.}2}_{-1{.}5}$& &$-49{.}7^{+1{.}1}_{-1{.}6}$	&\\
$\tau$ [Myr]   	&  $\sim5$& &$2{.}1^{+0{.}5}_{-0{.}5}$	&\\\bottomrule
\multicolumn{5}{p{10cm}}{$^\mathrm{a}$ Distance from dispersion measure for two models of the Galactic electron density \citep{1993ApJ...411..674T,2002astro.ph..7156C}. Proper motion from \citet{2003AJ....126.3090B}}\\
\multicolumn{5}{l}{$^\mathrm{b}$ \citet{2004ApJ...604..339C}}
\end{tabular}
}
\end{table*}

\begin{figure}
\centering
\includegraphics[width = 0.45\textwidth,viewport = 40 205 540 600]{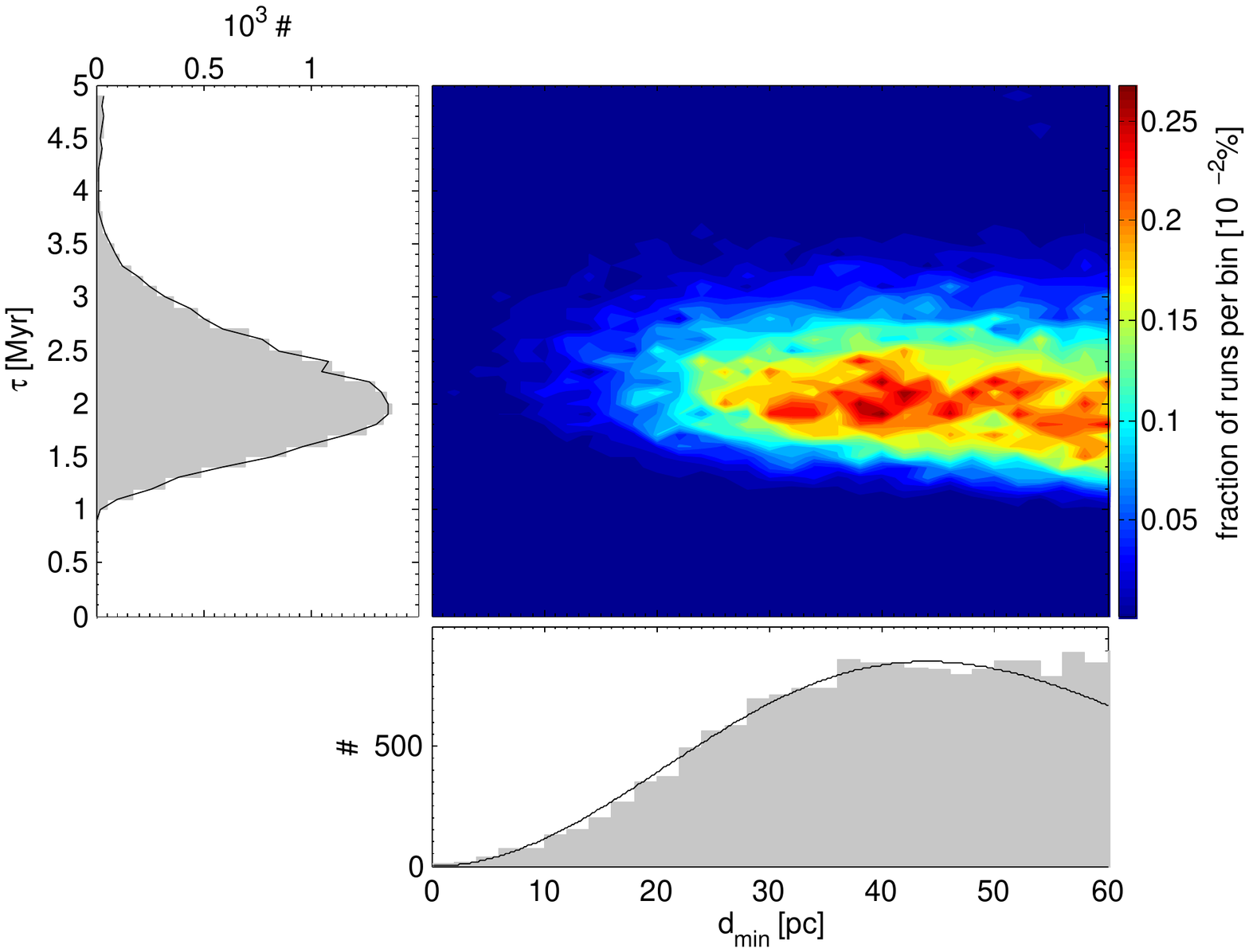}
\caption{Distributions of minimum separations $d_{min}$ and corresponding flight times $\tau$ for encounters between HIP 9470 and PSR\,J0152$-$1637. The solid curve drawn in the $d_{min}$ histogram (bottom panel) represents the theoretically expected distribution (see e.\,g. \citealt{2012PASA...29...98T}) with an expectation value for the separation of zero and a standard deviation of 31{.}1\,pc. The smallest separation between the two stars found after three million Monte Carlo runs was 1\,pc. This is expected from simulated test cases \citep{2013PhD...Nina}.}\label{fig:contour_hists}
\end{figure}

The spin-down age of PSR\,J0152$-$1637 of $\unit[10{.}1]{Myr}$ ($P$ and $\dot{P}$ from \citealt{2004MNRAS.353.1311H}) is a factor of five larger than our predicted kinematic age. For many neutron stars, kinematic ages are smaller than $\tau_{sd}$ by a factor of a few \citep[e.\,g.][]{2001A&A...365...49H,2011MNRAS.417..617T,2013MNRAS.435..879T}. The distance of the pulsar is rather uncertain and no parallax has been measured to date. Depending on the Galactic electron density model, dispersion measured distances range from about 500 to $\unit[800]{pc}$ \citep{1993ApJ...411..674T,2002astro.ph..7156C}. Our predicted distance of $833^{+167}_{-119}$ is in reasonable agreement with these estimates. \\
The distribution of separations $d_{min}$ between HIP 9470 and PSR\,J0152$-$1637 along with the flight time distribution is shown in \autoref{fig:contour_hists}. The solid line in the bottom panel represents the theoretical curve for the case that both objects met at the same position in space and is in good agreement with the $d_{min}$ histogram (see e.\,g. \citealt{2011MNRAS.417..617T}, \citealt{2013PhD...Nina} and \citealt{2014MNRAS.tmp..150T} for similar cases).\footnote{The theoretical curve shows the distribution of differences between two 3D Gaussians that are centred at the same point in space \citep[see e.\,g.][]{2001A&A...365...49H,2012PASA...29...98T}, i.\,e. from the distribution we assume that the two stars were at the same spatial position about $\unit[2]{Myr}$ ago.} \\
The encounter position cannot be associated with any particular stellar group that might have hosted the supernova, i.\,e., the predicted supernova event happened in isolation. 
This is possible if the neutron star progenitor itself was a runaway star. The former companion candidate HIP 9470 happens to be a binary star, as revealed by our spectroscopic monitoring. If PSR\,J0152$-$1637 and HIP 9470 were ejected during the same supernova event, this implies that the progenitor must have been at least a hierarchical triple system. This is not unlikely since stars ejected via gravitational interactions can be single, binary or triple stars \citep[e.g.][]{1989AJ.....98..217L}. The wide orbit of the binary might be a result of the supernova explosion.

\section{Conclusions}

We have carried out high-resolution spectroscopic monitoring of 30 stars previously included in the catalogue of candidate young runaway stars of \citet{2011MNRAS.410..190T}, for the purpose of measuring their radial velocities accurately. With this and other existing information we were able to calculate their Galactic orbits and investigate their origin in the context of runaway star scenarios. Ages for the targets were revised from previous estimates, and seven of the stars were found to be consistent with being younger than 100\,Myr, and therefore still plausible as runaway stars. The remaining 23 objects are old enough that we no longer consider them to be candidate runaway stars. We computed the past orbits of the seven younger stars, and for five of them we identified several young clusters or associations with known kinematics with which they appear to have crossed paths in the recent past, suggesting this may be their place of birth. Of these five stars, the properties of HIP 9470 are consistent with it having had a recent close encounter with the neutron star PSR J0152$-$1637, but not simultaneously with any of the clusters.  Seven of the 30 stars in the sample are found to be spectroscopic binaries, and orbital solutions for four of them are reported here.

\acknowledgements

NT and RN acknowledge partial support from DFG in the SFB/TR-7 Gravitational
Wave Astronomy.\\
This work has made use of BaSTI web tools (http://albione.oa-teramo.inaf.it/), the database by Lejeune and Schaerer (http://webast.ast.obs-mip.fr/equipe/stellar/, \citealt{2001A&A...366..538L}) and the Padova database of stellar evolutionary tracks and isochrones (http://pleiadi.pd.astro.it/).

\bibliographystyle{aa}
\bibliography{bib2}

\end{document}